\documentclass[aps,prb,twocolumn,superscriptaddress,showpacs,preprintnumbers,byrevtex,floatfix]{revtex4}

\usepackage{dcolumn}
\usepackage{bm}
\usepackage[usenames]{color}
\usepackage{amssymb}
\usepackage{amsfonts}
\usepackage{amsmath}
\usepackage[dvips]{graphicx}
\usepackage{graphpap}

\newcommand{\vac}{{\rm v}}

\begin{document}

\date{\today}
\title{Atomic defects and dopants in ternary Z-phase transition-metal nitrides CrMN with M=V, Nb, Ta investigated with density functional theory}

\author{Daniel F. Urban}
\email{daniel.urban@iwm.fraunhofer.de} 
\affiliation{Fraunhofer
Institute for Mechanics of Materials IWM, W\"ohlerstr. 11, 79108
Freiburg, Germany}

\author{Christian Els\"asser}
\affiliation{Fraunhofer Institute for Mechanics of Materials IWM,
W\"ohlerstr. 11, 79108 Freiburg, Germany}
\affiliation{University of Freiburg, Freiburg Materials Research Center (FMF), 
Stefan-Meier-Str. 21, 79104 Freiburg, Germany}

\begin{abstract}
A density functional theory study of atomic defects and dopants in ternary Z-phase transition-metal nitrides CrMN with M=V, Nb, or Ta is presented. Various defect formation energies of native point defects and of substitutional atoms of other metal elements which are abundant in the steel as well, are evaluated. The dependence thereof on the thermodynamic environment, i.e. the chemical conditions of a growing Z-phase precipitate, is studied and different growth scenarios are compared. The results obtained may help to relate results of experimental atomic-scale analysis, by atom probe tomography or transmission electron microscopy, to the theoretical modeling of the formation process of the Z phase from binary transition metal nitrides.
\end{abstract}


%
%

\maketitle

\section{Introduction}
In order to reduce the fuel consumption and CO$_2$ emission of fossil-fuel fired power plants, their thermal efficiency and therefore the steam inlet temperature should be raised as high as possible. This goal can only be achieved with the development of an appropriate material for  tubes and turbine rotors which is capable to withstand the fierce operational demands. During the past three decades sufficiently creep-resistant 9\% chromium steels were developed, which allowed to raise the steam temperature up to 615$^{\circ}$ C. These steels get their increased creep resistance mainly by the controlled precipitation of fine (V,Nb)N particles in the iron matrix.\cite{Maruyama2001,Taneike2003} A further raise of the steam temperature calls for steels with  even higher Cr contents in order to improve their corrosion and oxidation resistance. 
However, when increasing the Cr content of these ferritic-martensitic steels strengthened by fine (V,Nb)N particles to 11-12\% Cr, the precipitation of the thermodynamically stable Z phase\cite{Strang1996,Danielsen2006b}, CrMN (M=V,Nb,Ta), turned out to be unavoidable in long-time service.
Usually, these Z-phase particles are coarse and brittle, and they grow at the expense of the desired fine (V,Nb)N nitride particles.\cite{Hald2008,Chilukuru2009} Therefore, the Z-phase precipitation poses a severe problem being detrimental for the steel tubes in the power plant due to the decrease in creep rupture strength.\cite{Sawada2006,Sawada2008a,Hald2008,Golpayegani2008}
For a recent comprehensive review on the Z-phase precipitation in 9-12\% Cr steels see Ref. \onlinecite{Danielsen2016}.

An appealing concept to overcome the problems discussed above is provided by the idea to exploit the Z phase itself as a strengthening agent which is thermodynamically more stable than (V,Nb)N.\cite{Danielsen2009c} Hence the challenge is to control the precipitation of the Z phase in 12\% Cr martensitic creep resistant steels in such a way that fine particles are formed which are homogeneously dispersed and remain stable for long time. The cooperative research project Z-Ultra\cite{Zultra} was funded by the European Commission 
2013 -- 2016 to take this challenge. The atomic 
composition of chromium steels was varied in a series of test melts and the precipitation of 
nitrides was achieved by subsequent heat treatment.\cite{Liu2016} The materials of the test series were subjected to long-term creep-test experiments at various temperatures, and the microstructure of the materials was analyzed before and after mechanical loading.

Fundamental understanding of the Z phase, its crystal structure, its composition 
(stoichiometry), and its defects, is needed to relate results of experimental atomic-
scale analysis, by atom probe tomography (APT) or transmission electron microscopy (TEM), to 
theoretical modeling of the formation process of the Z phase from (V,Nb)N, and consequently to 
generate a feedback loop for the design and optimization of the steel composition and heat 
treatment. 

Experimentally, the Z phase was first discovered in austenitic steels containing niobium in the 1950's.\cite{Binder1950} However, it took several attempts\cite{Hughes1967,Ettmayer1971} before its crystal structure could finally be identified by Jack and Jack in 1972.\cite{Jack1972} Subsequently, mostly the extensive experimental work by Danielson and Hald \cite{Danielsen2006a,Danielsen2006b,Danielsen2007,Hald2008,Golpayegani2008,Danielsen2009a,Danielsen2009b,Cipolla2010a,Cipolla2010b,Danielsen2012,Danielsen2013} and by Sawada et al.\cite{Maruyama2001,Taneike2003,Sawada2006,Sawada2008a} contributed to the present understanding of the Z phase. The Z-phase particles which were examined showed large variations in composition which 
indicates the possibility of substituting a large number of atoms from the stoichiometric 
crystal structure by other, similar atoms. Already the early experiments\cite{Jack1972} identified minor concentrations of Fe and Mo and traces of other elements in the Z-phase precipitates. This was later confirmed by atom probe tomography measurements.\cite{Liu2010,Liu2016,Rashidi2017a}
While the perfect CrMN Z phase has a 1:1:1 ratio of Cr, transition-metal and nitrogen atoms, the experimentally determined composition varies. Moreover, in general the Z phase is of Cr(M,M')N type, with two transition metal species, e.g. V and Nb, sharing the same sublattice.\cite{Strang1996,Danielsen2006b,Danielsen2009a}  Understanding these observations calls for a detailed study of atomic point defects, mainly  substitutional defects and vacancies, in the Z phase crystal.

So far, recent theoretical studies using atomistic simulations focused on the thermodynamical stability and electronic structures of perfect Z-phase crystals, as well as on their elastic properties.\cite{Lazar2008,Kocer2009,Fors2011,Legut2012,Lv2014}

In this work we examine the Z-phase transition-metal nitrides CrMN with M=V, Nb, and Ta, their native atomic defects, and substitutional foreign atoms (impurities or dopants), by means of atomistic simulations using density functional theory. We hereby focus on the influences of deviations from the 1:1:1 stoichiometry and of foreign atoms on the energetics of Z-phase stability. The paper is organized as follows. Section \ref{sec:system} introduces the material system in detail and describes our theoretical approach. Our results are presented in Sec. \ref{sec:results}. First, results on structural and cohesive properties of perfect (defect-free) Z-phase crystals are given and compared to literature results in Sec. \ref{sec:structure}.
Native atomic defects in Z-phase crystals are studied in Sec. \ref{sec:native:defects}. Section \ref{sec:contact:MN} discusses the dependence thereof on the chemical environment, i.e. the growth conditions of the Z-phase precipitate.
The incorporation of other metal species from the steel matrix into the Z phase is studied in Sec. \ref{sec:other:defects}, followed by separate discussions on the role of nitrogen (Sec. \ref{sec:N:vacancy}) and carbon (Sec. \ref{sec:carbon}). Interactions between defects which can potentially favor certain off-stoichiometries and lead to defect-clustering effects are discussed in Sec. \ref{sec:interaction}.
We summarize and conclude our work in Sec. \ref{sec:summary}.

\section{Material system and theoretical approach}
\label{sec:system}

\subsection{Crystal structure of Z phases}
The Z phase has a tetragonal crystal structure, its space group is P4/nmm  (\# 129), and it 
consists of six atoms in the tetragonal unit cell, see Fig. 1 (left). 
All the atoms of the 
three species in the tetragonal unit cell are located at Wyckoff positions (2c) which are 
(0,0,$z$) and (0.5,0.5,-$z$), yielding the three internal structural parameters, $z_{\rm M}$, $z_{\rm Cr}$ and $z_{\rm N}$.
For the investigation of atomic defects in this work we have used a 
4x4x2 supercell consisting of 192 atoms, see Fig. 1 (right). This figure illustrates well the layered 
structure of the Z phase, which is composed of alternating pairs of Cr and MN layers. The 
individual MN-layer and Cr-layer pairs can be imagined as each two (001)-oriented atomic layers of bulk MN with B1 (face-centered cubic rocksalt) structure and bulk Cr with A2 (body-centered-cubic tungsten) structure, respectively.

\begin{figure}[]
      \begin{center}
      \includegraphics[width=2.5cm]{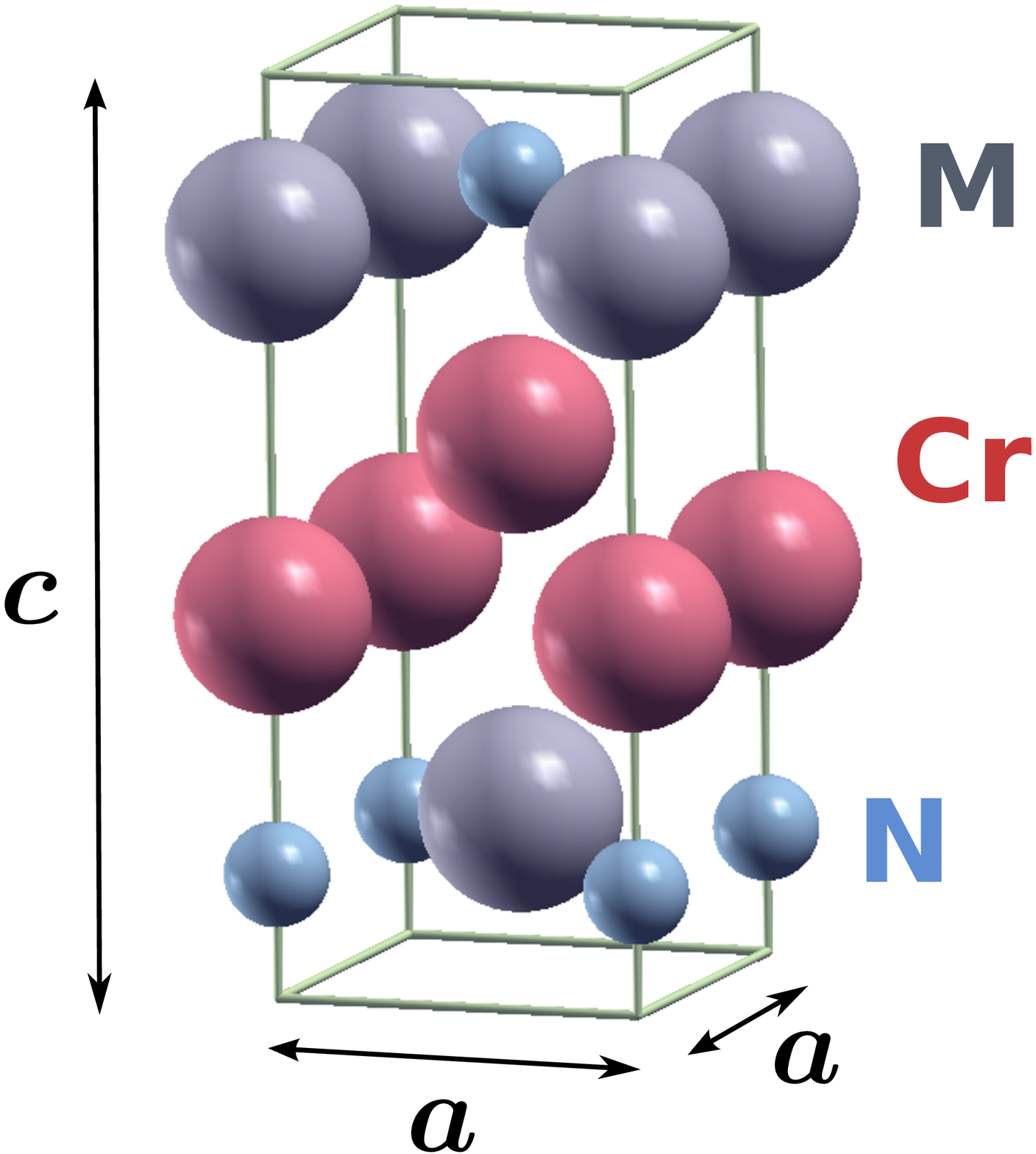}
      \includegraphics[width=5.5cm]{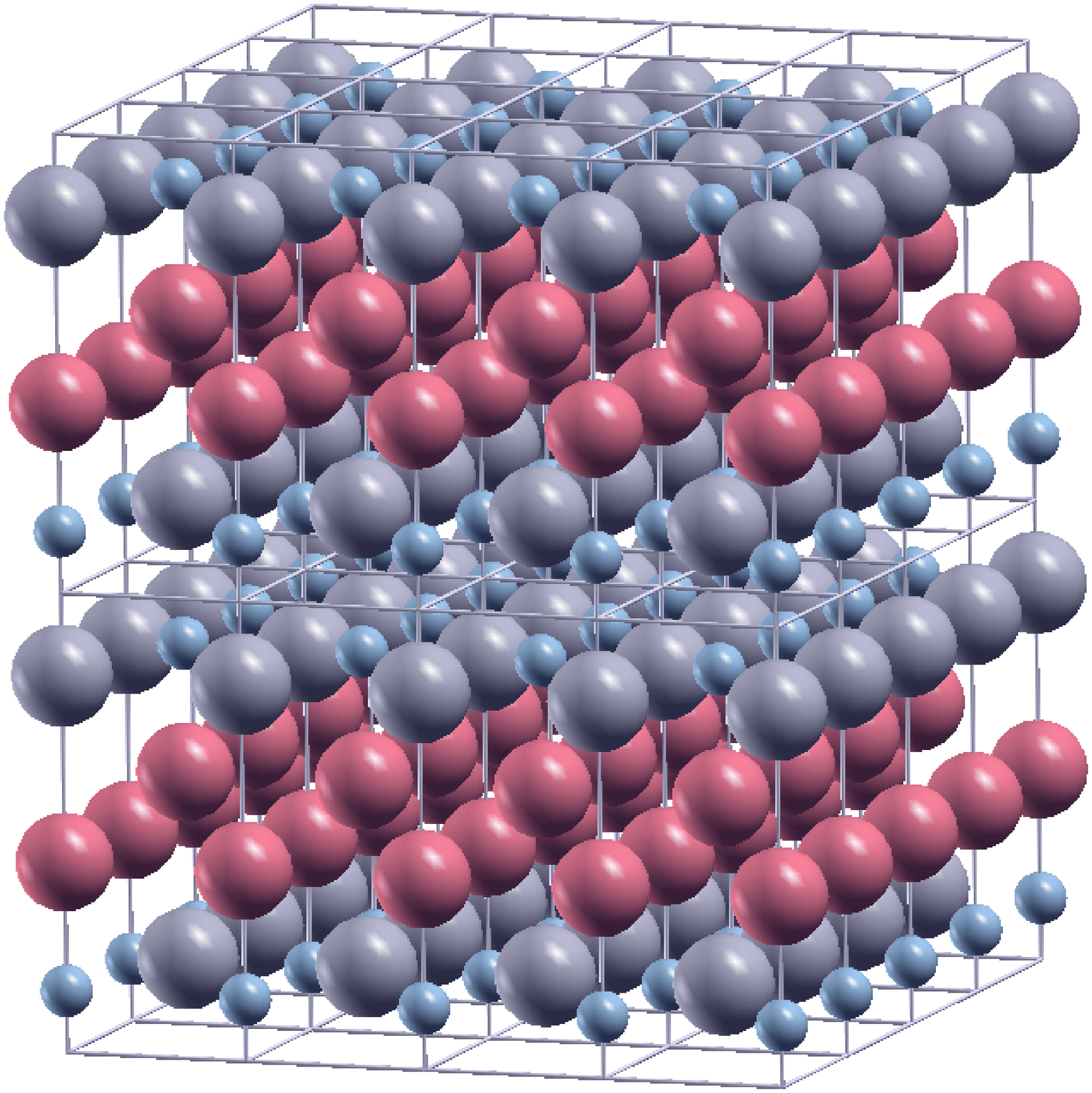}
\caption{(Color online)
Z-phase structure with an alternating sequence of two bcc-like layers of Cr atoms 
(red) and two rocksalt-like nitride layers, where V/Nb/Ta atoms are shown in gray 
and N atoms are shown in blue. The 192 atoms supercell (right) used for the study of 
point defects consists of 4x4x2 primitive cells (left) of six atoms, each.
\label{fig:geometry}}
\end{center}
\end{figure}

\subsection{Computational settings for DFT calculations}
\label{sec:comp:setting}
All DFT calculations were carried out using the projector augmented-wave (PAW) method\cite{bl94} as implemented in the VASP code.\cite{kr96,kr99} The PBE-GGA was used for the exchange-correlation functional, and PAW pseudopotentials with 5, 13, 13, 11, and 14 valence electrons for N, V, Nb, Ta, and Cr, respectively. All calculations were carried out with a plane-wave cutoff energy of 600 eV, a $16\times 16\times 6$ Monkhorst-Pack k-mesh for the 6-atom unit cell, and a Gaussian broadening of 0.1 eV. Accordingly, a $4\times 4\times 3$ Monkhorst-Pack k-mesh was used for the supercells consisting of 4x4x2 tetragonal unit cells.

For the calculation of defect formation energies, 
the defect-containing supercells were structurally relaxed by optimizing the atomic positions to zero forces by means of the Broyden-Fletcher-Goldfarb-Shanno (BFGS) algorithm. 
The lattice parameters (and thus the supercell volume) were fixed to the equilibrium values of the defect-free bulk.
We have verified that the supercells are large enough to compensate the local volume change due to the presence of the point defect and that the overall stress is reasonably small.

\subsection{Defect formation energies and chemical potentials}

The formation energy of a substitutional defect with one atom X substituting an atom Y is calculated as the difference in total energies,
\begin{equation}
\label{eq:Eform:gen}
\Delta E_f[{\rm X}_{\rm Y}] = 
E_{\rm tot}[{\rm CrMN}\!:\!{\rm X}_{\rm Y}]-E_{\rm tot}[{\rm CrMN}]-\mu_{\rm X}+\mu_{\rm Y},
\end{equation}
where $\mu_{X/Y}$ denotes the chemical potential of element X/Y. 
An analogous formula is used for the calculation of the formation energy of a vacancy $\vac_Y$ on sublattice Y by formally setting $\mu_{\vac}\equiv0$. The chemical potentials $\mu$ of all involved elements in general depend on the full chemical environment and thereby on the presence of competing phases.

A well-defined limiting case is to use the most stable elemental phases as references in order to evaluate the chemical potentials $\mu^{(0)}$ of the atoms involved in the defect formation. Therefore, e.g., the total energies per atom of the pure elemental bcc metals V, Nb, Ta, and Cr were used as the chemical potentials $\mu^{(0)}$ of the respective metals. For nitrogen, the gas phase of N$_2$ molecules\cite{footnote1} is used as reference to obtain $\mu_{\rm N}^{(0)}$. Correspondingly, the defect formation energy with respect to the elemental phases is then given by
\begin{equation}
\label{eq:Eform:0}
\Delta E_f^{(0)}[{\rm X}_{\rm Y}] = 
E_{\rm tot}[{\rm CrMN}\!:\!{\rm X}_{\rm Y}]-E_{\rm tot}[{\rm CrMN}]-\mu_{\rm X}^{(0)}+\mu_{\rm Y}^{(0)}.
\end{equation}
In the following, the relative chemical potentials 
\begin{equation}
	\Delta\mu_X=\mu_X-\mu^{(0)}_X
\end{equation} 
are used. The formation energy (\ref{eq:Eform:gen}) of a substitutional defect with one atom X substituting an atom Y is then related to $\Delta E_f^{(0)}[{\rm X}_{\rm Y}]$ by
\begin{equation}
\label{eq:Eform:deltaMu}
\Delta E_f[{\rm X}_{\rm Y}] = \Delta E_f^{(0)}[{\rm X}_{\rm Y}]-\Delta\mu_{\rm X}+\Delta\mu_{\rm Y}.
\end{equation}

\subsection{Formation enthalpies and equilibria of competing phases}
\label{sec:ternaryChemPot}

\begin{figure}[]
\begin{center}
     \includegraphics[width=0.9\columnwidth]{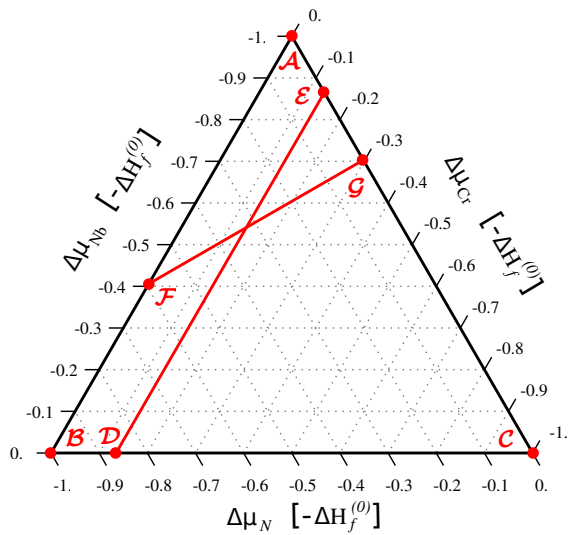}
\caption{(Color online) Range of possible values for the relative chemical potentials $\Delta\mu_{\rm Cr}$, $\Delta\mu_{\rm Nb}$, and $\Delta\mu_{\rm N}$ under the boundary conditions (\ref{eq:mu:cond:CrMN}) and (\ref{eq:mu:cond:boundary}), for the case of the CrNbN Z-phase, given in units of $-\Delta H_f^{(0)}$[CrNbN]. The points $\mathcal{A}$--$\mathcal{G}$ and the respective connecting red lines mark special limiting cases which are discussed in detail in the text.
\label{fig:ternaryChemPot}}
\end{center}
\end{figure}

In equilibrium, the Z phase is stable if
\begin{equation}
\label{eq:mu:cond:CrMN}
\Delta H_f^{(0)}[{\rm CrMN}] = \Delta\mu_{\rm Cr} + \Delta\mu_{\rm M} + \Delta\mu_{\rm N},
\end{equation}
where $\Delta H_f^{(0)}$[CrMN] is the formation enthalpy\cite{footnote2} of CrMN from its elemental constituents. Note that $\Delta H_f^{(0)}<0$. Furthermore,
the upper boundary for the chemical potentials is given by 
\begin{equation}
\label{eq:mu:cond:boundary}
\Delta\mu_X\leq0,
\end{equation}
i.e. the threshold for precipitation of the elemental phase of component X.
The range of possible values for $\{\Delta\mu_{\rm Cr},\Delta\mu_{\rm M},\Delta\mu_{\rm N}\}$ under the conditions (\ref{eq:mu:cond:CrMN}) and (\ref{eq:mu:cond:boundary}) can be illustrated in a diagram analogous to a ternary phase diagram. Figure \ref{fig:ternaryChemPot} exemplifies such a diagram for the specific case of CrNbN. (The respective diagrams for CrVN and CrTaN only differ in the energy scale $\Delta H_f^{(0)}[{\rm CrMN}]$ and the exact position of the points $\mathcal{D}$--$\mathcal{G}$, which are slightly shifted with respect to the points plotted in the figure.)
The three corners of the triangle, points $\mathcal{A}$, $\mathcal{B}$, and $\mathcal{C}$ in Fig. \ref{fig:ternaryChemPot}, correspond to one $\Delta\mu_X$ taking its smallest (i.e. most negative) value $\Delta\mu_{\rm X}=\Delta H_f^{(0)}$ whereas the other two relative chemical potentials are zero. These three limiting cases are referred to as
($\mathcal{A}$) the \emph{M-poor} limit, ($\mathcal{B}$) the \emph{N-poor} limit, and 
($\mathcal{C}$) the \emph{Cr-poor} limit. The three borders of the triangle, lines $\overline{\mathcal{B}\mathcal{C}}$, $\overline{\mathcal{A}\mathcal{C}}$, and $\overline{\mathcal{A}\mathcal{B}}$, correspond to one $\Delta\mu_X$ being zero and are referred to as the \emph{M-rich}, \emph{N-rich}, and  \emph{Cr-rich} limits, respectively.

The mostly observed and currently widely accepted Z-phase formation process is the transformation of MN nitride particles by gradual interdiffusion of Cr atoms.\cite{Danielsen2009b,Cipolla2010a}
In this scenario, the growing Z-phase particle is in direct thermodynamic contact with the remaining parent MN nitride particle which sets the boundary condition
\begin{eqnarray}
\label{eq:mu:cond:MN}
\Delta H_f^{(0)}[{\rm MN}] &\leq& \Delta\mu_{\rm M} + \Delta\mu_{\rm N}
\end{eqnarray}
to the chemical potentials involved.
Here, equality corresponds to the phase equilibrium which together with Eq. (\ref{eq:mu:cond:CrMN}) leaves only one parameter undetermined. Correspondingly, the condition of phase equilibrium between CrMN Z-phase and MN is represented by the line $\overline{\mathcal{D}\mathcal{E}}$ in Fig. \ref{fig:ternaryChemPot} where the point $\mathcal{D}$ ($\mathcal{E}$)
corresponds to $\Delta\mu_{\rm N}$ ($\Delta\mu_{\rm M}$) taking the value of the ratio of the enthalpies of formation $\Delta H_f^{(0)}$[MN]/$\Delta H_f^{(0)}$[CrMN].
The region of instability of the MN nitride, determined by the inequality (\ref{eq:mu:cond:MN}), is given by the area left of the line $\overline{\mathcal{D}\mathcal{E}}$.

Recently, an alternative nucleation and growth mechanism in which Cr$_2$N precipitates are gradually transformed has been proposed to occur under certain conditions.\cite{Sawada2006,Lee2015,Kim2017}
In this case, the chemical potentials are related by Eq. (\ref{eq:mu:cond:CrMN}) and
\begin{eqnarray}
\label{eq:mu:cond:Cr2N}
\Delta H_f^{(0)}[{\rm Cr}_2{\rm N}] &\leq& 2\Delta\mu_{\rm Cr} + \Delta\mu_{\rm N}.
\end{eqnarray}
The condition of phase equilibrium between CrMN Z-phase and Cr$_2$N is represented by the line $\overline{\mathcal{F}\mathcal{G}}$ in Fig. \ref{fig:ternaryChemPot}
and the region of instability of the Cr$_2$N nitride, determined by inequality (\ref{eq:mu:cond:Cr2N}), is given by the area below this line.

\section{Results}
\label{sec:results}

\begin{table*}[t]
\begin{tabular}{l l c c c c c c c}
\hline\hline	 				 		 			 
 & &  $a$ [\AA]& $c$ [\AA] & $z_{\rm M}$ & $z_{\rm N}$ & $z_{\rm Cr}$ & 
 $\Delta H_f^{(0)}$ [eV] & $\Delta H_f^{\rm (prec)}$ [eV]\\
\hline
CrVN & & & & & & & & \\	 	 	 	 	 	 	 	 	 	 
\hline	 				 		 			 
this work				& DFT (VASP, PBE, 600eV, 16x16x6)	& 
2,860 & 7,130 & 0,834 & 0,136 & 0,409  
& -2,652 & -0,294\\
Fors  (2011)\cite{Fors2011}		& DFT (VASP, PBE, 700eV, 12x12x6) 	& 
2,862	& 7,148	& 0,834	& 0,136	& 0,409  
&&\\
Lazar (2008)\cite{Lazar2008}	& DFT (VASP, PW91)		& 
2,860	& 7,140 & 0,838	& 0,140	& 0,409  
& -2,334 &\\
Kocer (2009)\cite{Kocer2009}	& DFT (VASP, PW91, 408eV, 8x8x4)	& 
2,857	& 7,125	& 		&		&				
& -2,260 &\\	
Legut (2012)\cite{Legut2012}	& DFT (VASP, PBE, 600eV, 35x35x14)	& 
2,86	& 7,134	& 	 	&		&				
& -2,263	& -0,279\\ 
Danielsen (2006)\cite{Danielsen2006a}	& Experiment					& 
2,86	& 7,39	&		&		&	& & \\	
\hline					
CrNbN & & & & & & & & \\
\hline				 		 			 
this work				& DFT (VASP, PBE, 600eV, 16x16x6)	& 
3,023	& 7,364	& 0,824	& 0,139	& 0,414	 
& -2,736	& -0,367\\
Lv (2014)\cite{Lv2014}			& DFT (CASTEP, PBE, 400eV, 8x8x3)	& 
3,026	& 7,389	& 0,824	& 0,139	& 0,414  
& -2,462 &\\
Fors (2011)\cite{Fors2011} 		& DFT (VASP, PBE, 700eV, 12x12x6)	& 
3,039	& 7,401	& 0,823	& 0,140	& 0,415	 
&&\\
Legut (2012)\cite{Legut2012}	& DFT (VASP, PBE, 600eV, 33x33x14)	& 
3,021	& 7,367	&		&		&			
& -2,181	& -0,351 \\	
Kocer (2009)\cite{Kocer2009}	& DFT (VASP, PW91, 408eV, 8x8x4)	& 
3,041 & 7,387 &		&		&				
& -2,191 &\\
Jack  (1972)\cite{Jack1972} 	& Experiment						& 
3,037	& 7,391	&		&			&		& &\\
Ettmayer (1971)\cite{Ettmayer1971} 	& Experiment 					& 
3,029	& 7,360	& 0,825	& 0,125	& 0,410	  & &\\
\hline
CrTaN & & & & & & & & \\
\hline	 				 		 			 
this work 				& DFT (VASP, PBE, 600eV, 16x16x6)	& 
3,019	& 7,390	& 0,821	& 0,135	& 0,414	 
& -2,691	& -0,620\\
Fors (2011)\cite{Fors2011}		& DFT (VASP, PBE, 700eV, 12x12x6)	& 
3,017	& 7,407	& 0,821 & 0,135 & 0,414	 &&\\ 
Ettmayer (1971)\cite{Ettmayer1971} 		& Experiment				& 
3,026 & 7,390	&		&		&		&&	\\
Danielsen (2009)\cite{Danielsen2009a}	& Experiment				& 
2,96	& 7,39 	&		&		&		&&	\\	
\hline\hline
\end{tabular}
\caption{Compilation of lattice parameters (a,c), internal coordinates (z$_{\rm M}$, z$_{\rm N}$, z$_{\rm Cr}$), and  enthalpies of formation for the bulk Z-phase crystals CrMN (M=V, Nb, Ta) obtained by DFT simulations or determined by experiments. The internal coordinates refer to the Wyckoff positions (0, 0, z) and (0.5, 0.5, -z). Enthalpies of formation are either with respect to the elemental phases ($\Delta H_f^{(0)}$) or the transformation of MN precipitates ($\Delta H_f^{\rm (prec)}$) and are given per formula unit.
For the listed theoretical results taken from literature, the employed DFT code, exchange-correlation functional, cut-off energy, and k-mesh are given as well for better comparison. 
\label{tab:struc:param}
}
\end{table*}

\subsection{Structural and cohesive properties of perfect Z-phase crystals}
\label{sec:structure}
We have calculated the equilibrium cohesive and structural properties of the bulk Z-phase crystals by 
fitting the Murnaghan equation of state to a dataset of minimal total energies calculated for 
different unit-cell volumes. Optimization of the lattice-parameter ratio c/a and 
relaxation of the atomic positions were achieved by minimizing the elastic stress and the forces acting on the atoms, respectively, in each step, i.e. for each unit-cell volume. By this  procedure, the equilibrium values for the lattice parameters and atomic positions were obtained. 
In table \ref{tab:struc:param} our results are compiled and compared to those of other authors as found in literature. Altogether the agreement is reasonably good. Deviations among  different theoretical studies can most likely be attributed to the variations in the numbers of valence electrons utilized in the pseudo potential of each element and the convergence parameters used, such as cutoff energy and k-point mesh.
Only few experimentally obtained values are available, and for those references that report structural parameters, it is not obvious, whether the samples had the perfect 1:1:1 stoichiometry. 

For calculating the enthalpy of formation we have followed two different approaches for which the results are also given in table \ref{tab:struc:param}. The first one corresponds to the formation from \emph{elemental phases}, 
\begin{equation}
\label{eq:deltaH:0}
\Delta H_f^{(0)} = 
E_{\rm tot}[{\rm CrMN}]-\mu_M^{(0)}-\mu_{\rm Cr}^{(0)}-\mu_{{\rm N}}^{(0)},
\end{equation}
where the $\mu^{(0)}_{\rm M/Cr/N}$ are the chemical potentials of the respective atoms in their most stable elemental phases, namely M and Cr in the bcc crystal phase and N$_2$ gas.\cite{footnote1}
Note that the lower values of $\Delta H_f^{(0)}$ compared to other theoretical predictions originate mainly in our choice of $\mu_{{\rm N}}^{(0)}$ which includes a gas correction term  $\sim 0.35$eV adding on the pure GGA result.\cite{footnote1}
The second possibility to evaluate the enthalpy of formation corresponds to the formation from the B1-rocksalt transition metal nitride \emph{precipitate phases} and bulk bcc Cr,
\begin{equation}
\label{eq:deltaH:prec}
\Delta H_f^{\rm (prec)} = 
E_{\rm tot}[{\rm CrMN}]-E_{\rm tot}[{\rm MN(B1)}]-\mu_{\rm Cr}^{(0)}.
\end{equation}
The latter is the more relevant case when discussing the Z-phase formation process which is considered to occur by the transformation of nitride particles via Cr interdiffusion.\cite{Danielsen2009b,Cipolla2010a}
For all three Z phases both formation enthalpies are found to be significantly negative which reflects on the one hand the strong driving force for the growth of the Z-phase precipitates and on the other hand the superior thermodynamic stability compared to the MN nitrides.

For the evaluation of $\Delta H_f^{\rm (prec)}$ it may be even more suitable to consider the chemical potential of Cr atoms dissolved in the iron matrix as appropriate reference. We found the latter to be 0.08 eV lower in the dilute limit compared to $\mu_{\rm Cr}^{(0)}$, see App. \ref{app:contact:aFe}. This in turn shifts the values of the corresponding $\Delta H_f^{\rm (prec)}$ to $-0.214$/$-0.287$/$-0.540$ eV/f.u. for M=V, Nb and Ta, respectively.

\subsection{Atomic defects in Z-phase crystals}
\label{sec:native:defects}

For the three Z-phase compounds CrMN (M=V,Nb,Ta) we have calculated formation energies of atomic defects, thereby focusing on vacancies and substitutional defects.
The results for $\Delta E_f^{(0)}[{\rm X}_{\rm Y}]$, cf. Eq. (\ref{eq:Eform:0}), obtained for the native point defects of the four metals (X, Y = Cr, V, Nb, Ta) are compiled in Tab. \ref{tab:native:defects}.
The formation energies of atomic defects are related to their concentrations. 
The concentration of atomic defects generated by thermal fluctuations at a given temperature T is proportional to $\exp(-E_f/k_BT)$, where $k_B$ is Boltzmann's constant. Note that $k_BT\sim 0.03$eV at room temperature whereas $k_BT\sim0.09$eV for T=\mbox{700$^\circ$ C}. The following conclusions are drawn from the results obtained under metal-rich/N-poor conditions:
\begin{enumerate}
\item
The formation of metal vacancies in Z-phase crystals is energetically not favorable. 
\item
A structural disorder, i.e. an interchange of Cr and M atoms on their respective lattice sites (anti-site defects) while maintaining the exact stoichiometry is energetically also not favorable.
\item
Formation of mixed Z-phases, i.e., replacing host M atoms in CrMN by other M' atoms (M, M' = V, Nb, Ta), requires only small energies ($<0.3$eV) in most cases or is even energetically favorable in some cases. This is in accordance with experimental observations of mixed Cr(V,Nb)N and Cr(Nb,Ta)N phases.\cite{Strang1996,Danielsen2006b,Cipolla2010a,Danielsen2009a,Danielsen2009b}
\item
Variation of the stoichiometry by replacing Cr atoms in CrMN by host M atoms or other M' atoms, (M, M' = V, Nb, Ta) is connected with a low cost in energy for the Z phases CrNbN and CrTaN, or it is even energetically favorable for the Z phase CrVN. 
\end{enumerate}

\begin{table}[]
\begin{tabular}{l c c c}
\hline\hline
 & CrVN & CrNbN & CrTaN  \\
\hline
$\vac_{\rm M}$	& 2.61 & 2.26 & 1.60 \\
$\vac_{\rm Cr}$ & 2.76 & 2.58 & 2.33 \\
\hline
V$_{\rm M}$ &	 ---  & 0.27 &	0.13 \\
Nb$_{\rm M}$&	0.14  & ---	 & -0.07 \\
Ta$_{\rm M}$&	0.10  &	0.02 &	---  \\
Cr$_{\rm M}$&	1.37  &	1.79 &	1.67 \\
\hline
V$_{\rm Cr}$ & -0.22 & -0.49 &-0.50 \\
Nb$_{\rm Cr}$&	0.84 &	0.10 & 0.15 \\
Ta$_{\rm Cr}$&	0.82 &	0.08 & 0.17 \\
\hline\hline
\end{tabular}
\caption{
Formation energies $\Delta E_f^{(0)}[{\rm X}_{\rm Y}]$ of atomic point defects in the Z-phase crystals CrMN (M=V,Nb,Ta), with respect to the most stable metal elemental phases. All values are given in eV.  
\label{tab:native:defects}
}
\end{table}

\subsection{Z-phase precipitate in thermodynamic contact with MN or Cr$_2$N nitride particles}
\label{sec:contact:MN}

The metal-rich environment discussed in the previous section actually is a limiting case, cf. discussion in Sec. \ref{sec:ternaryChemPot} and Fig. \ref{fig:ternaryChemPot}.
In general, the chemical potentials $\mu$ of all involved elements depend on the full chemical environment and thereby on the presence of competing phases. This section compares the defect formation energies for the two different Z-phase growth scenarios, namely the 
gradual transformation of (i) MN nitride particles or (ii) Cr$_2$N precipitates.
Note that here it is not relevant to consider Laves phases MCr$_2$ since there is no experimental evidence for Z-phase growth by nitriding Laves phase precipitates.

\begin{figure}[]
\begin{center}
     \includegraphics[width=0.9\columnwidth]{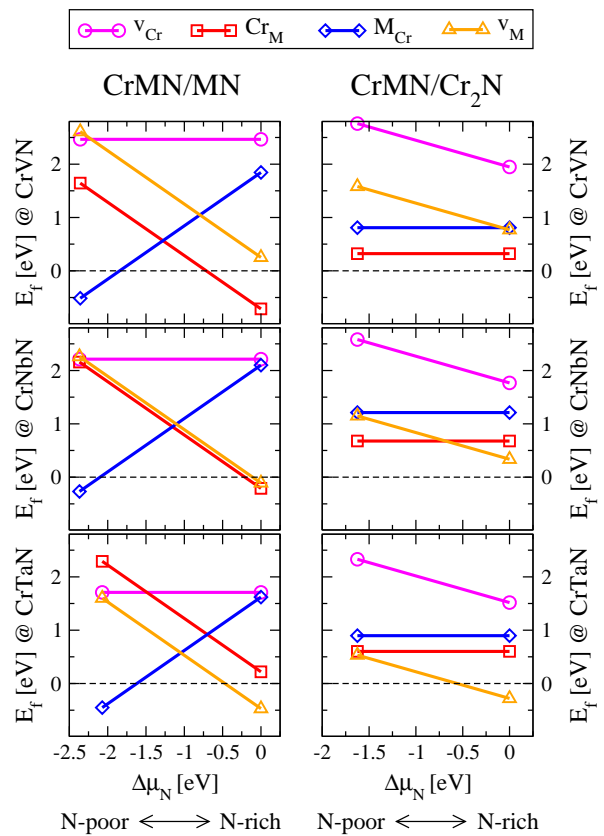}
\caption{(Color online) Variation of defect formation energies of native point defects 
(v$_{\rm Cr}$, Cr$_{\rm M}$, v$_{\rm M}$, and M$_{\rm Cr}$) in the CrMN Z-phase in thermodynamic contact with MN (left column) or Cr$_2$N (right column) precipitates, as function of the relative chemical potential $\Delta\mu_{\rm N}$ of nitrogen. The variation in relative chemical potentials (x-axis) corresponds to the variation along the lines $\overline{\mathcal{D}\mathcal{E}}$ and $\overline{\mathcal{F}\mathcal{G}}$ of Fig. \ref{fig:ternaryChemPot}, respectively
\label{fig:Eform:Fkt:muN}}
\end{center}
\end{figure}

In case (i), the growing Z-phase particle is in thermodynamic contact with the host MN nitride particle\cite{Danielsen2009b,Cipolla2010a} which yields the two conditions (\ref{eq:mu:cond:CrMN}) and (\ref{eq:mu:cond:MN}). Equilibrium between CrMN and MN determines two out of three relative chemical potentials $\{\Delta\mu_{\rm Cr},\Delta\mu_{\rm M},\Delta\mu_{\rm N}\}$. This leaves only one parameter undetermined that we choose to be $\Delta\mu_{\rm N}$ which relates to the nitrogen concentration. 
Evaluating the formation energies (\ref{eq:Eform:deltaMu}) for the defects introduced in Sec. \ref{sec:native:defects} under these boundary conditions leads to the results shown in Fig. \ref{fig:Eform:Fkt:muN}, left column, for CrVN (top), CrNbN (middle), and CrTaN (bottom). The variation in $\Delta\mu_{\rm N}$ corresponds to the variation along the line $\overline{\mathcal{D}\mathcal{E}}$ of Fig. \ref{fig:ternaryChemPot}. The N-rich limit $\Delta\mu_{\rm N}=0$ refers to point $\mathcal{E}$, in accordance with the discussion of Sec. \ref{sec:ternaryChemPot}.

On the other hand, if the Z phase grows from a Cr$_2$N precipitate,\cite{Sawada2006,Lee2015,Kim2017} the boundary conditions (\ref{eq:mu:cond:CrMN}) and (\ref{eq:mu:cond:Cr2N}) yield the variation of the relative chemical potentials along the line 
$\overline{\mathcal{F}\mathcal{G}}$ of Fig. \ref{fig:ternaryChemPot} and point $\mathcal{G}$ refers to the N-rich case $\Delta\mu_{\rm N}=0$.
Correspondingly, the defect formation energies evaluated under these boundary conditions lead to the results shown in the right column of Fig. \ref{fig:Eform:Fkt:muN}. 

Comparing the results of the two columns of Fig. \ref{fig:Eform:Fkt:muN} shows that 
for both scenarios of Z-phase growth, Cr vacancies are not relevant. Moreover, a cross-over of the relative occurrence probability between M vacancies and Cr atoms substituting M is observed, when moving from M=V to Nb and Ta. A significant difference between the two growth mechanisms is observed though with respect to the expected deviations from the perfect 1:1:1 ratio of CrMN. When the Z-phase growth is hosted by a MN particle, M$_{\rm Cr}$ defects are always energetically favorable in the N-poor case, whereas for N-rich conditions, Cr$_{\rm M}$ is favorable for M=V, Cr$_{\rm M}$ and v$_{\rm M}$ are almost degenerate for M=Nb, and v$_{\rm M}$ is favorable for M=Ta. On the other hand, when the Z-phase grows from Cr$_2$N particles, substoichiometry is only predicted for the N-rich environment of CrTaN.

A variation in the M:Cr ratio is also frequently observed in experiments, depending on the specific precipitate studied and the history of the sample. For Cr(Nb,V)N Z-phase precipitates, the (Nb,V):(Cr,Fe) ratio ranges from 50:50\cite{Danielsen2006b} to 42:48\cite{Cipolla2010a}. Note that in both studies 5 at.\% Fe is found in the Z-phase which is attributed to Fe substituting on the Cr lattice sites (see the following Sec. \ref{sec:other:defects}). Liu \emph{et al.}\cite{Liu2016} report a (Ta,Nb,V):(Cr,Fe) ratio of 43:56 for Z-phase precipitates with 39 at.\% Ta. 

A recent experimental study using atom probe tomography however reports a Ta:Cr ratio of 38:62 (based on atomic percentage) for Ta-based Z-phase precipitates.\cite{Rashidi2017a} These results are in line with the early experiment by Ettmayer\cite{Ettmayer1971}, who postulates that this large excess in Cr is caused by a significant amount of Ta being substituted by Cr.
This strong deviation from the ideal 1:1 ratio cannot straightforwardly be explained by our results for isolated point defects. From Fig. \ref{fig:Eform:Fkt:muN} we can infer that only the N-rich growth condition may lead to a Ta:Cr ratio significantly smaller than 1:1 via a substantial concentration of Ta vacancies. However, defect-defect interactions need to be taken into account for large defect concentrations far from the dilute limit (see Sec. \ref{sec:interaction}). Although the defect formation energy for Cr$_{\rm Ta}$ substitutional defects becomes rather low under N-rich conditions it is always higher than the energy of Ta vacancies.

\subsection{Other metal species from a steel matrix substituted in the Z phases}
\label{sec:other:defects}

\begin{figure}[]
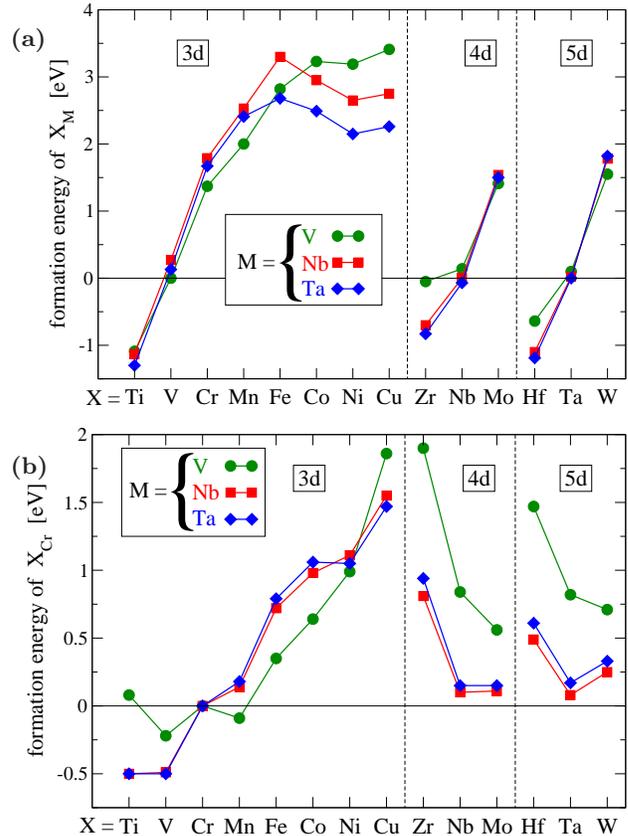

\begin{center}
  \setlength{\unitlength}{1mm}
  \begin{picture}(85,108)(0,0)
     \put(0,105){\bf (a)}
     \put(0,48){\bf (b)}
     \put(5,57){\includegraphics[width=77mm,draft=false]{Fig4a__dataXsubM_v2.eps}}
     \put(2,0){\includegraphics[width=80mm,draft=false]{Fig4b__dataXsubCr_v2.eps}}
  \end{picture}
\caption{(Color online)
Compilation of the data from Tabs. \ref{tab:native:defects} and \ref{tab:other:defects}  showing the general trend observed for
the formation energies $E_f^{(0)}$ of substitutional defects (a) on host metal sites, X$_{\rm M}$, and (b) on Cr sites, X$_{\rm Cr}$, evaluated for M-rich/Cr-rich conditions. 
\label{fig:Eform:trends}}
\end{center}
\end{figure}

Besides the transition-metal elements Cr, V, Nb, and Ta considered above, many further chemical elements are present in the iron matrix of steel as a result of the challenging task to obtain optimum compositions with respect to application-specific target criteria for steels. Therefore it is also a relevant and interesting aspect to take into account a potential transfer of some other elemental species from the steel matrix to the Z-phase precipitates. Namely the elements Mn, Co, Ni, Cu, Mo, W, and of course Fe itself, too, are considered as potential candidates to form substitutional defects in the Z phases. For completeness we have also considered Ti, Zr, and Hf, although they have so far not been considered as alloying element for the Z-phase strengthened steels. 
We have calculated the atomic defect formation energies for these metallic species substituting on either Cr or M  sites in the Z phases CrMN using the same 192-atom supercells as discussed in Sec. \ref{sec:native:defects}.  The defect formation energies were calculated with reference to pure metals and the respective most stable phases with bcc, fcc, or hcp structures were taken for calculating their chemical potentials. The results are summarized in Tab. \ref{tab:other:defects}, and Fig. \ref{fig:Eform:trends} displays the general trend. 
This specific choice of reference phases for the chemical potentials again reflects the 
\emph{M-rich} and \emph{Cr-rich} limiting cases, which correspond to $\Delta\mu_{\rm M}=0$ and $\Delta\mu_{\rm Cr}=0$, respectively. 
The following conclusions concerning the substitution of foreign metal elements into the Z phase are drawn:
\begin{enumerate}

\item
The substitutional incorporation of metallic species which are often present in steels, namely Mn, Fe, Co, Ni, Cu, Mo, or W, on host-metal M sites in CrMN is found to be energetically unfavorable and therefore not likely to happen.

\item Besides the intermixing of V, Nb, and Ta on the host-metal M sites (discussed in Secs. \ref{sec:native:defects} and \ref{sec:contact:MN}), the substitution of M by Ti, Zr, and Hf is connected with an energy gain. Therefore,
these elements are likely to be incorporated into the Z phase in considerable amounts (if available in the steel).

\item The incorporation of all considered element species on Cr sites in CrMN is always connected with a lower cost in energy as the respective substitution on the M sites. Depending on the Z phase under consideration, the substitution of Cr atoms by Ti, Mn, Fe, Mo, or W atoms is accompanied by such a low cost in energy or even an energy gain, that considerable non-zero concentrations of such defects are expected. 

\end{enumerate}

Experimentally, often a Fe content of roughly \mbox{5 at. \%} (with respect to the metal content) is reported for Z-phase particles.\cite{Strang1996,Danielsen2006b,Cipolla2010a,Danielsen2009a,Danielsen2009b}
In addition, Strang and Vodarek\cite{Strang1996} report 0.7--0.9 at.\% of Mo, while the more refined analysis by atom probe tomography (APT) yields traces of other metals as well.\cite{Liu2016}
Despite its obvious advantages in terms of precision in distinguishing different elements, the APT method requires a normalization with respect to the steel matrix, for determining the composition of a (generally) irregularly shaped precipitate.\cite{Rashidi2017b} Specifically, the signal of an unknown amount of surrounding steel matrix needs to be subtracted from the acquired raw data. Usually this is done via the choice of an element which is not expected to be incorporated in the precipitate but which still shows a statistically significant signal. The theoretical prediction of the hierarchy of defect formation energies presented above may guide this post-processing of raw APT data. 

Commonly the incorporation of Fe atoms is assumed to occur on Cr lattice sites, while Mo is claimed to substitute for V or Nb in the Z phase Cr(V,Nb)N.\cite{Danielsen2006a} While our results confirm the higher probability of Fe$_{\rm Cr}$ compared to Fe$_{\rm M}$ for all three M, the evaluated defect formation energies also favor Mo to substitute for Cr, especially in a Z phase containing Nb or Ta. 

\begin{table}[]
\begin{tabular}{l c c c}
\hline\hline
 & CrVN & CrNbN & CrTaN  \\
\hline
Ti$_{\rm M}$	&-1.09 &-1.13 &-1.30 \\
Mn$_{\rm M}$	& 2.00 & 2.53 &	2.41 \\
Fe$_{\rm M}$	& 2.82 & 3.30 &	2.68 \\
Co$_{\rm M}$	& 3.23 & 2.95 &	2.49 \\
Ni$_{\rm M}$	& 3.19 & 2.65 &	2.15 \\
Cu$_{\rm M}$	& 3.41 & 2.75 &	2.26 \\
Zr$_{\rm M}$	&-0.05 &-0.70 &-0.83 \\
Mo$_{\rm M}$	& 1.41 & 1.54 &	1.50 \\
Hf$_{\rm M}$	&-0.64 &-1.10 &-1.19 \\
W$_{\rm M}$	    & 1.55 & 1.78 &	1.82 \\
\hline
Ti$_{\rm Cr}$	& 0.08 &-0.50 &-0.50 \\
Mn$_{\rm Cr}$	&-0.09 & 0.14 &	0.18 \\
Fe$_{\rm Cr}$	& 0.35 & 0.72 &	0.79 \\
Co$_{\rm Cr}$	& 0.64 & 0.98 &	1.06 \\
Ni$_{\rm Cr}$	& 0.99 & 1.11 &	1.05 \\
Cu$_{\rm Cr}$	& 1.86 & 1.55 &	1.47 \\
Zr$_{\rm Cr}$	& 1.90 & 0.81 & 0.94 \\
Mo$_{\rm Cr}$	& 0.56 & 0.11 &	0.15 \\
Hf$_{\rm Cr}$	& 1.47 & 0.49 & 0.61 \\
W$_{\rm Cr}$	& 0.71 & 0.25 &	0.33 \\
\hline\hline
\end{tabular}
\caption{
Calculated formation energies $E_f^{(0)}$ (given in eV) of substitutional atomic defects on M and Cr sites in the Z-phase crystal structures of CrMN (M=V, Nb, Ta). 
\label{tab:other:defects}
}
\end{table}

\subsection{Nitrogen vacancies in the Z phases}
\label{sec:N:vacancy}

So far we have considered point defects on the two metal sublattices of the Z phases. Concerning the nitrogen sublattice, there are two interesting defects to study, namely the nitrogen vacancy and the substitution of N by carbon which is also abundant in the steel (see next subsection). 

For the calculation of the formation energy of nitrogen vacancies in the Z phase the choice of the correct reference has a strong influence on the magnitude of the obtained values. We have made and compared four different choices, (i) the gas phase of N$_2$ molecules,\cite{footnote1} (ii) N atoms dissolved on interstitial (octahedral) sites in the bcc iron matrix of a ferritic steel,  (iii) N in bulk binary transition-metal nitrides MN of rocksalt structure in N-poor conditions,
and (iv) N in bulk CrMN Z-phase crystal in N-poor conditions.

The choice (i) is rather academic than realistic for properties of defects in precipitates buried inside steels. The choice (ii) is well suited and guided by the knowledge that nitrogen atoms occupy octahedral interstitial sites in the ferritic (bcc) iron matrix. \cite{Jack1950,Christofaro1977}
Compared to the gas phase, we obtain a relative chemical potential of $\Delta\mu_{\rm N}=-0.14$eV in the dilute limit (see App. A).

The reference choice (iii) for obtaining the chemical potential of N is guided by the experimental observation that Z-phase particles form in the steel matrix by a gradual transformation of binary transition-metal-nitride particles MN of rocksalt (B1) crystal structure.\cite{Danielsen2009a,Cipolla2010a}
For metal-rich/nitrogen-poor conditions ($\Delta\mu_{\rm M}=0$) the relative chemical potential of nitrogen is therefore given by $\Delta\mu_{\rm N} = \Delta H_f^{(0)}[{\rm MN(B1)}]$, cf. point $\mathcal{D}$ in Fig. \ref{fig:ternaryChemPot}.
Finally, reference choice (iv) corresponds to the situation when the binary MN nitride particle has been fully transformed to Z phase and this precipitate continues to grow by consuming Cr and M from the iron matrix. The limiting N-poor/M-rich/Cr-rich environment corresponds to  
$\Delta\mu_{\rm M}=0$, $\Delta\mu_{\rm Cr}=0$, and the constrained minimum $\Delta\mu_{\rm N}=\Delta H_f^{(0)}[{\rm CrMN}]$, cf. point $\mathcal{B}$ in Fig. \ref{fig:ternaryChemPot}.

The inspection of the results compiled in table \ref{tab:N:vacancy} yields the following conclusions. Under N-rich conditions the formation of nitrogen vacancies in the Z-phase crystal is energetically suppressed. During the first stage of Z-phase growth, the available nitrogen from the matrix and the host MN precipitate is being consumed, and nitrogen vacancies become more and more likely when approaching the N-poor regime. The formation of nitrogen vacancies is favored even more in the final stage of the Z-phase precipitation, when the particle growth is fed via particle diffusion through the steel matrix and 
the N-poor regime is reached.

\begin{table}[]
\begin{tabular}{l c c c c}
\hline\hline
reference for N reservoir & CrVN & CrNbN & CrTaN  \\
\hline
N$_2$ gas 	& 2.27 & 2.89 & 2.93 \\
N dissolved in bcc-Fe 
			& 2.13 & 2.75 & 2.79 \\
N-poor MN (B1) 
			& -0.09 & 0.52 & 0.86 \\
N-poor CrMN 
			& -0.38 & 0.15 & 0.24 \\ 
\hline\hline
\end{tabular}
\caption{Defect formation energies of nitrogen vacancies ($\vac_N$) in Z phases CrMN, calculated for three choices of the reference phase (reservoir) for the chemical potential of nitrogen. Energies are given in eV.
\label{tab:N:vacancy}
}
\end{table}

\begin{table}[]
\begin{tabular}{l c c c}
\hline\hline
reference for N/C reservoirs & CrVN & CrNbN & CrTaN  \\
\hline
graphite and N$_2$
			& 1.86 & 1.70 & 1.47 \\
M-rich MC/MN (B1) 
			& 0.36 & 0.43 & 0.56 \\
\hline\hline
\end{tabular}
\caption{Defect formation energies of carbon substituting nitrogen (C$_N$) in Z phases CrMN, calculated for two choices of the reference phase (reservoir) for the chemical potentials $\mu_{\rm C}$ and $\mu_{\rm N}$. Energies are given in eV.
\label{tab:C_sub_N}
}
\end{table}

\subsection{Carbon incorporation in the Z phases}
\label{sec:carbon}

Since carbon is an important and abundant element in steels, we also consider the nitrogen of the Z phase to be partially replaced by C atoms, similar to what is known from the mixed M(N,C) phases. Also here, the judicious choice of reference is crucial for the evaluation of defect formation energies. For the evaluation of the formation energy of the substitutional C$_{\rm N}$ defect we again distinguish two different cases: (i) the most stable elemental phases graphite and N$_2$ gas and
(ii) the exchange of C and N with bulk binary transition-metal carbides MC / nitrides MN of rocksalt structure in M-rich conditions (see Table \ref{tab:C_sub_N}). 
As in the previous section, the latter choice is more suitable for describing the experimental situation during the growth of the Z-phase precipitate by consumption of a M(N,C) host particle. The defect formation energies in the latter case are such that small concentrations of C within the Z phase are to be expected. Recent measurements by atom probe tomography report 0.5--1.3 at. \% carbon content, depending on the Z-phase precipitate examined.\cite{Rashidi2017a}

\subsection{Interaction between defects}
\label{sec:interaction}

The interaction between atomic point defects may sometimes lower the energy cost for their formation and thereby favor lattice imperfections via defect-defect clustering. We have considered the pairing of a selection of point defects discussed in the previous sections in order to estimate the magnitude of this effect in the Z-phase crystal structure. The interaction energy between two atomic point defects ${\rm X}_{\rm Y}$ and ${\rm X'}_{\rm Y'}$ is calculated via the difference in total energies 
\begin{eqnarray}
E_{int} &=& E_{tot}[{\rm CrMN}\!:\!({\rm X}_{\rm Y}+{\rm X'}_{\rm Y'})]+ E_{tot}[{\rm CrMN}]
\nonumber\\
\label{eq:Eint}
&&-E_{tot}[{\rm CrMN}\!:\!{\rm X}_{\rm Y}]-E_{tot}[{\rm CrMN}\!:\!{\rm X'}_{\rm Y'}].
\end{eqnarray}
Here, $E_{tot}[{\rm CrMN}\!:\!({\rm X}_{\rm Y}+{\rm X'}_{\rm Y'})]$ is calculated using a structure model where the two point defects are neighboring each other. In most cases, there are several different possible orientations for the defect pair. Hence we took the configuration with the lowest energy. 
Note that the interaction energy as defined in Eq. (\ref{eq:Eint}) does not depend on the choice of reference chemical potentials for the elemental phases.

\begin{table}[]
\begin{tabular}{l | c c c c | c c c c|}
\hline\hline
&CrVN &&&&CrTaN&&&\\	
			& $\vac_{\rm M}$ & v$_{\rm N}$ & M$_{\rm Cr}$ & Cr$_{\rm M}$  
			& $\vac_{\rm M}$ & v$_{\rm N}$ & M$_{\rm Cr}$ & Cr$_{\rm M}$\\
\hline
$\vac_{\rm M}$	& 0.43       &  0.08     &   0.01      & -0.12   
                & 0.85       &  0.37     &  -0.11      & -0.12  \\ 
$\vac_{\rm N}$	&            &  0.10     &  -0.01      & -0.18  
                &            & -0.04     &   0.06      & -0.30  \\
M$_{\rm Cr}$	&            &           &   0.01      & -0.04  
            	&            &           &  -0.04      & -0.13  \\
Cr$_{\rm M}$	&            &           &             & -0.05  
                &            &           &             & -0.06  \\
\hline\hline
\end{tabular}
\caption{Interaction energies of pairs of neighboring point defects in CrVN and in CrTaN Z-phase. Energies are given in eV.
\label{tab:interaction}
}
\end{table}

The interaction energies for the intrinsic point defects in the Z phases CrVN and CrTaN  are listed in Tab. \ref{tab:interaction}. The corresponding results for CrNbN (not shown) are very similar to those for \mbox{CrTaN}. The general conclusions which can be drawn from this compilation of data are the following:
\begin{enumerate}
\item There is no pronounced tendency for defect-defect-clustering.
\item Substitutional Cr energetically prefers to be neighbor to a N vacancy or, to less extent, to a host-metal vacancy.
\item The formation of pairs of substitutional Cr does not require additional energy (instead there even is a small energy gain). This eventually may favor the clustering of Cr$_{\rm M}$ under M-poor chemical environment.
\item Host-metal vacancies repel each other significantly. 
\end{enumerate}
Therefore, significant deviations from the 1:1 ratio of Cr:M in the Z-phase towards a surplus of Cr will not occur via a larger concentration of host-metal vacancies. On the other hand, the clustering of Cr$_{\rm M}$ is energetically possible. 

\section{Summary and conclusions}
\label{sec:summary}

We have studied atomic defects and dopants in ternary Z-phase transition-metal nitrides CrMN (M=V, Nb, or Ta) within the framework of density functional theory. Defect formation energies of native point defects, i.e. mutual substitutions or vacancies of V, Nb, Ta, and Cr, as well as of the substitution by other metal atoms which are present in the steel, namely Ti, Mn, Fe, Co, Ni, Cu, Zr, Mo, Hf, and W, have been evaluated. The influence of the chemical potentials of the elements involved in the formation of a substitutional point defect or a vacancy was exemplified for various thermodynamic conditions which represent different chemical environments of a growing Z-phase precipitate.

Our results confirm the experimental observation\cite{Strang1996,Danielsen2006b,Cipolla2010a,Danielsen2009a,Danielsen2009b} that an interchange 
of the host M atoms in CrMN by other M' atoms (M, M' = V, Nb, Ta) is likely for ambivalent temperatures which leads to mixed Cr(M,M')N phases when M and M' both are available in the steel. 

Small non-stoichiometries Cr$_{1\pm x}$M$_{1\mp x}$N are expected for the specific chemical environment of a MN nitride particle hosting the Z-phase precipitate-growth. N-poor/M-rich conditions will yield a surplus of M, while N-rich/M-poor conditions result in a surplus of Cr.
This is in agreement with the reported experimentally observed compositions of Z-phase precipitates which tend to favor a slight Cr surplus of a few atomic percent.
For the alternative growth scenario in which Cr$_2$N particles are gradually transformed to Z phase, non-stoichiometry is only expected for CrTaN in N-rich growth conditions.
The significant non-stoichiometry of the experimentally observed\cite{Rashidi2017a,Ettmayer1971} Cr$_{1.2}$Ta$_{0.8}$N cannot straight forwardly be derived from our results for isolated point defects. Nevertheless, the evaluation of defect-defect interaction energies hints towards the possibility of clustering Cr$_{\rm Ta}$ defects as a result of a small energy gain.

The analysis of the substitutional incorporation of other metal atoms which are present in the steel shows that Ti, Mn, Fe, Mo, or W atoms may substitute for Cr because of a low cost in energy or even an energy gain, that considerable non-zero concentrations of these elements are expected in the Z phase. The substitution of these elements on the host metal sublattice is connected with significantly larger defect formation energies which yields negligible defect concentrations in comparison. On the other hand, the incorporation of Co, Ni, and Cu is energetically suppressed. This may support the post-processing analysis of APT data which allow for distinguishing the signal of the Z-phase from the surrounding steel matrix.\cite{Rashidi2017b} 

\section{Acknowledgments} 
Financial support for this work was provided by the European Commission through contract No. 309916, project Z-ultra. We thank Hermann Riedel, Hans-Olof Andren, and Fang Liu for many valuable discussions.

\appendix

\section{Chemical potential of Cr and N in bcc Fe} 
\label{app:contact:aFe} 

\begin{figure}[t]
\begin{center}
   \includegraphics[width=\columnwidth]{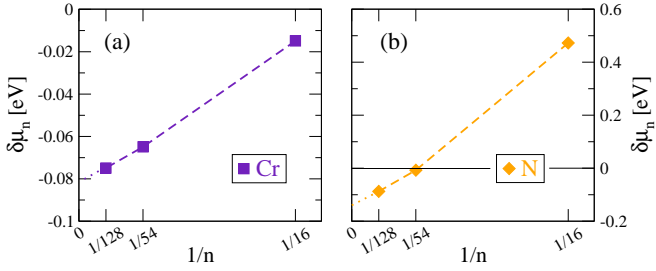}
\caption{(Color online)
Calculated chemical potentials of (a) individual Cr atoms dissolved on substitutional sites in a bcc Fe crystal and (b) individual N atoms located on octahedral interstitial sites as function of the supercell size (number of atoms n) which is used for the calculation.
\label{fig:chem:pot:Cr:and:N}}
\end{center}
\end{figure}  

\begin{table}[t]
\vspace*{0.5cm}
\begin{tabular}{l c c c}
\hline\hline
nitride & a [\AA] & c/a & $\Delta H_f^{(0)}$ [eV/f.u.] \\
\hline
 VN  & 2.914  & 1.0 & -2.359 \\
 NbN & 3.125  & 1.0 & -2.368 \\
 TaN & 3.128  & 1.0 & -2.071 \\
 VC  & 2.940  & 1.0 & -0.860 \\
 NbC & 3.170  & 1.0 & -1.105 \\
 TaC & 3.169  & 1.0 & -1.164 \\
 Cr$_2$N & 4.771  & 0.922 & -1.626 \\
\hline
\end{tabular}
\caption{Lattice parameter a, aspect ratio c/a, and enthalpies of formation $\Delta H_f^{(0)}$ of binary nitride structures used as reference for the calculation of chemical potentials.
\label{tab:binary:nitrides}
}
\end{table}

The chemical potential of Cr dissolved in a ferritic iron matrix phase can be obtained by considering an individual Cr atom substituting a Fe atom in a n-atom supercell of a bcc Fe crystal. With the energy difference
\begin{eqnarray}
\delta \mu_n^{\rm (Cr)}& = &
E_{tot}[{\rm Fe}_{n-1}{\rm Cr}]-(n-1)\mu^{(0)}_{\rm Fe}- \mu_{\rm Cr}^{(0)}\quad
\end{eqnarray}
we evaluate $\Delta\mu[{\rm Cr\;in\;Fe}_{\rm bcc}] = \lim_{n\rightarrow\infty}\delta \mu_n^{\rm (Cr)}$ in the dilute limit ($n\rightarrow\infty$).
We have used supercells with n=16, 54, and 128 atoms. The k-point density was adjusted accordingly by using 8x8x8, 5x5x5, and 4x4x4 Monkhorst-Pack k-meshes respectively, and we have conducted spin polarized calculations. The result is displayed in Fig. \ref{fig:chem:pot:Cr:and:N} (a) and indicates a very good linear scaling behavior for $\delta E_n$ versus $1/n$. Calculations with a supercell of 54 atoms already yield sufficiently accurate results. 

Correspondingly, we evaluate the chemical potential of N dissolved in a ferritic iron matrix phase.
Since N is known to occupy octahedral interstitial sites in the bcc iron crystal we define
\begin{eqnarray}
\delta \mu_n^{\rm (N)} &=& 
E_{tot}[{\rm Fe}_{n}{\rm N}]-n\mu_{\rm Fe}^{(0)} - \mu_{\rm N}^{(0)}.
\end{eqnarray}
The results, which are displayed in Fig. \ref{fig:chem:pot:Cr:and:N} (b), again show a very good linear scaling behavior. Yet there is a much stronger dependency on supercell size, as compared to the results of Fig. \ref{fig:chem:pot:Cr:and:N} (a) for the substitutional Cr because N as an interstitial point defect causes a stronger distortion of the surrounding Fe atoms.

\section{Reference binary nitrides}
\label{app:reference}
For the evaluation of the relative chemical potentials under Z-phase precipitate growth conditions, cf. Eqs. (\ref{eq:mu:cond:MN}) and (\ref{eq:mu:cond:Cr2N}), the formation enthalpies of the MN nitrides and MC carbides in rocksalt crystal structure (B1) and the trigonal Cr$_2$N are required. The corresponding space groups are Fm$\overline{3}$m (\#225) for
the B1 structure and P$\overline{3}$1m (\#162) for Cr$_2$N.
Optimization procedure and computational settings are the same as those described in Sec. \ref{sec:comp:setting}. For the chromium nitride phases, spin polarized calculations have been conducted. The results are summarized in Tab. \ref{tab:binary:nitrides}. 


\begin{thebibliography}{42}
\bibitem[{foo({\natexlab{b}})}]{footnote2}
\bibinfo{howpublished}{All phase formation enthalpies given in this work are
  calculated at zero pressure and temperature.}
\end{thebibliography}

\end{document}